\documentclass{amsart}
 
\usepackage{amsmath}
\usepackage{amsfonts}
\usepackage{stmaryrd}
\usepackage{epsfig}
\usepackage{graphics} 

\usepackage{pgfplots,pgfplotstable}
\usepgfplotslibrary{groupplots}
\usetikzlibrary{shapes,arrows,snakes,calendar,matrix,backgrounds,folding,calc,positioning,spy}
\usepackage{tikz}
\usepackage{booktabs}

\usepackage[utf8]{inputenc}
\usepackage[T1]{fontenc}
\usepackage{graphicx}
\usepackage{amssymb,amsmath}
\usepackage{url}
\usepackage{lastpage}

\usepackage{hyperref}
\definecolor{darkblue}{rgb}{0,0,.7}
\hypersetup{colorlinks=true,linkcolor=darkblue,citecolor=darkblue}

\usepackage{todonotes}

\newcommand\numberthis{\addtocounter{equation}{1}\tag{\theequation}}

\def\doubleunderline#1{\underline{\underline{#1}}}

\newcommand{\tr}{\operatorname{tr}}
\newcommand{\dev}{\operatorname{dev}}
\renewcommand{\div}{\operatorname{div}}

\title[High-order projection-based upwind method for ILES]{High-order projection-based upwind method for implicit large eddy simulation}

\author[P.~L.~Lederer]{Philip L. Lederer}
\address{Wiedner Hauptstraße 8-10, 1040 Vienna, Austria}
\email{philip.lederer@tuwien.ac.at}

\author[X.~Mooslechner]{Xaver Mooslechner}
\address{Wiedner Hauptstraße 8-10, 1040 Vienna, Austria}\email{xaver.mooslechner@tuwien.ac.at}

\author[J. Sch\"oberl]{Joachim Sch\"oberl}
\address{Wiedner Hauptstraße 8-10, 1040 Vienna, Austria}
\email{joachim.sch\"oberl@tuwien.ac.at}

\usepackage[backend=bibtex,doi=false,isbn=false, url=false,
eprint=false, firstinits=true,
style=numeric,defernumbers=false,sorting=none,style=numeric-comp,maxbibnames=99]{biblatex}
\begin{document}

\begin{abstract}
We assess the ability of three different approaches based on
high-order discontinuous Galerkin methods to simulate under-resolved
turbulent flows. The capabilities of the mass conserving mixed stress
method as structure resolving large eddy simulation solver are
examined. A comparison of a variational multiscale model to no-model
or an implicit model approach is presented via numerical results. In
addition, we present a novel approach for turbulent modeling in
wall-bounded flows. This new technique provides a more accurate
representation of the actual subgrid scales in the near wall region
and gives promising results for highly under-resolved flow problems.
In this paper, the turbulent channel flow and periodic hill flow
problem are considered as benchmarks for our simulations. 
\end{abstract}
 

\maketitle

\section{Introduction}\label{intro}

\subsection{Motivation}

A complete capture of the broad range of time and length scales
presented in high Reynolds number flows can be obtained by direct
numerical simulation (DNS). This highly computational demanding
approach for simulating turbulent flows requires to resolve the
smallest appearing scales, which leads to exceptionally large problems
to solve. The most common alternative to overcome this prohibitively
expensive technique is the large eddy simulation (LES) \cite{sagaut}.
In LES, only the large scale structures of the flow are resolved and
the fine scale part is modeled using an explicit subgrid scale (SGS)
model. By now, LES simulation is a well-established field in
computational engineering and numerous different SGS models have been
developed till today. A further refined method for scale resolving
turbulent flows is the variational multiscale simulation (VMS)
\cite{VMShughes}. By separating the turbulent flow regime in large and
small resolved scales, the VMS approach allows to treat them
differently.

The introduction of discontinuous Galerkin (DG) discretization methods
has gained notable attention for simulating complex turbulent flows. A
significant amount of research has been spent in the field of DG
methods for LES
\cite{ILES_peraire2,Bolemann2015,collis,plata,assessment,zhang,ferrer,bassi}.
Due to the low dissipation (dissipation on high-order terms only),
high robustness and flexibility, DG approaches in general prove to be
very suitable for the multiscale characteristic of LES. These
conditions have lead to the development of no-model or implicit LES
(ILES). While conventional LES methods rely heavily on the accuracy of
SGS models, ILES or also known as under-resolved DNS uses its
intrinsic dissipation mechanism of the discretization itself as model.
Therefore, by no explicit model incorporated in the simulation, the
ILES has several advantages and has been successfully used in
\cite{ILES_christoph,ILES_peraire2,gassner2013accuracy,uranga,moura}.

In this work, we are interested in a comparison of VMS to ILES for
different wall-bounded turbulent flow scenarios. Our main goal is to
introduce and analyze an entirely new ILES approach, which further
improves the in-built LES capabilities of the underlying
discretization. This more enhanced ILES method uses a projection-based
upwind term as the convection stabilization.

We consider the mass conserving mixed stress method (MCS) introduced
in \cite{lederer} for solving the Navier-Stokes equations. This method
can be seen as a mixed formulation of the exactly divergence-free
$H(\mathrm{div})$-conforming hybrid discontinuous Galerkin (HDG)
method from \cite{lehrenfeld}. The advantage is that MCS results in
less (the minimal) numerical dissipation needed to guarantee inf-sup
stability of the underlying Stokes system as compared to an interior
penalty HDG approach, which makes it very attractive as baseline
discretization. The numerical results have been obtained by using the
open-source finite element framework NGSolve/Netgen \cite{ngsolve}.

\subsection{Overview}

The rest of the paper is organized as follows. We introduce the
spatial and temporal discretization in section \ref{discretization}. A
derivation of the different modeling approaches is given in section
\ref{methods}. We consider the turbulent channel flow problem in
section \ref{channel} and the periodic hill flow problem in section
\ref{periodichill}. The mesh convergence and accuracy of the methods
for both flow problems is investigated respectively. The final part,
section \ref{summary}, summarizes the observations made in this work
and gives conclusions.

\subsection{Mathematical Model}

Let $\Omega \subset \mathbb{R}^{3}$ be a connected and bounded
Lipschitz domain and $[0,T_{e}]$ a given time interval with end time
$T_e \in\mathbb{R}^+$. We consider the velocity field $\underline{u}$
and the pressure $p$ as the solution of the incompressible
Navier-Stokes equations
	\begin{subequations} \label{NSE}
		\begin{alignat}{2}
	\nabla\,\cdot\,\underline{u} &=0  && \qquad\textnormal{in}\:\Omega\times[0,T_{e}], \label{NSE_C} \\
	\frac{\partial\underline{u}}{\partial t} + (\underline{u}\,\cdot\,\nabla)\underline{u}
	 - 2\nu\nabla\,\cdot\,\epsilon(\underline{u}) + \nabla p &=  \underline{f} && \qquad \textnormal{in}\:\Omega\times[0,T_{e}], \label{NSE_M} \\
	\underline{u} &=\underline{u}_{0} && \qquad \textnormal{in}\:\Omega, t=0, \label{NSE_I}\\
	\underline{u} &=0 &&\qquad \textnormal{in}\:\partial\Omega_{D}\times[0,T_{e}], \label{NSE_B}
\end{alignat}
\end{subequations}
where $\underline{u}_{0}$ is a compatible initial condition. The
boundary is split into two part such that $\partial \Omega = \partial
\Omega_D \cup \partial \Omega_P$. On $\partial\Omega_{D}$ we prescribe
a zero no-slip (Dirichlet) value $\underline{u}=0$, and on
$\partial\Omega_{P}$ we consider periodic boundary conditions. In
addition we denote by $\underline{f}$ a (sufficiently smooth) external
force, and by $\nu\in \mathbb{R}^+\setminus \{ 0 \}$ the constant
kinematic viscosity. In \eqref{NSE_M}, the term $\epsilon
(\underline{u}) = \frac{1}{2}(\nabla\underline{u} +
\nabla\underline{u}^T)$ is the (symmetric) strain rate tensor, and
$\nabla\,\cdot\,\epsilon(\underline{u})$ indicates the divergence
applied to each row of $\epsilon(\underline{u})$.


\section{Solver for turbulent incompressible flows}\label{discretization}

\subsection{Notation}

Let $\mathcal{T}_h$ be a triangulation of the domain $\Omega$ into
hexahedral elements. The skeleton $\mathcal{F}_h$ denotes the set of
all element facets which is split into two sets $\mathcal{F}_h =
\mathcal{F}^I_h \cup \mathcal{F}^D_h$ where $\mathcal{F}^I_h$ denotes
the set of all interior and periodic facets and $\mathcal{F}^D_h$
indicates the facets at the Dirichlet boundary. 
Now let $\underline{\phi}$ be an element-wise
smooth vector-valued function. For arbitrary elements $T, T' \in
\mathcal{T}_h$ with a common facet $E \in \mathcal{F}_h$ let
$\underline{\phi}^T = \underline{\phi}|_T$ and $\underline{\phi}^{T'}
= \underline{\phi}|_{T'}$. We define the jump
$\llbracket\cdot\rrbracket$ and average $\langle\cdot\rangle$
operators by 
\begin{align*}
\llbracket\underline{\phi}\rrbracket = \underline{\phi}^T|_E - \underline{\phi}^{T'}|_E, \quad \textrm{and} \quad \langle\underline{\phi}\rangle = \frac{1}{2}(\underline{\phi}^T|_E + \underline{\phi}^{T'}|_E). 
\end{align*}
On the element boundary $\partial T$ the tangential trace operator is
given by  
\begin{equation*}
\underline{\phi}_t = \underline{\phi}-(\underline{\phi}\cdot\underline{n})\underline{n},
\end{equation*}
where $\underline n$ is the outward pointing normal vector.
Additionally, we introduce the normal-normal and normal-tangential
trace operators for smooth matrix-valued functions
$\doubleunderline{\phi}$ by
\begin{align*}
 \doubleunderline{\phi}_{nn} = (\doubleunderline{\phi} n)\cdot n, \quad \textrm{and} \quad \doubleunderline{\phi}_{nt} = \doubleunderline{\phi}n-(\doubleunderline{\phi}_{nn})n.
\end{align*}
The $\mathrm{skew}(\cdot)$ operator and the deviatoric part of a matrix is given as
\begin{equation*}
\mathrm{skew}(\doubleunderline{\phi})=\frac{1}{2} ( \doubleunderline{\phi} - \doubleunderline{\phi}^T ), \quad \textrm{and} \quad 
\dev({\doubleunderline{\phi}}) =
\doubleunderline{\phi} - \frac{1}{3} \tr(\doubleunderline{\phi}) I,
\end{equation*}
where $\tr(\cdot)$ is the matrix trace and $I$ is the identity matrix.

In this work we include the range into the notation of the spaces.
Thus while $H^1(\Omega, \mathbb{R})$ denotes the standard Sobolev
space of order one, $H^1(\Omega, \mathbb{R}^3)$ denotes its
vector-valued version. 
The Sobolev space $H(\mathrm{div})$ is defined as
\begin{equation*}
H(\mathrm{div},\Omega)=\{\underline{v}\in L^2(\Omega,\mathbb{R}^3):\nabla\,\cdot\,\underline{v}\in L^2(\Omega,\mathbb{R})\}.
\end{equation*}
Note that for all the above Sobolev spaces, we use a zero subscript to
denote that the corresponding continuous trace vanishes on $\partial
\Omega_D$, i.e. $H_0(\div, \Omega)$ denotes all functions $v \in
H(\div, \Omega)$ such that $v_n = 0$ on $\partial \Omega_D$. For
conforming (i.e. normal continuous) approximations of functions in
$H(\mathrm{div},\Omega)$, we consider the Raviart-Thomas finite element
space $RT^k(\mathcal{T}_h)$, see \cite{raviart, boffi2013mixed}, where
$k$ is a given order. In addition we denote by
$\mathbb{P}^{k}(\mathcal{T}_h, \mathbb{R})$  the space of all
element-wise polynomials defined on $\mathcal{T}_h$ whose total order
is less or equal $k$. 

\subsection{The discrete model - the MCS method}\label{mcs}

In this work, we consider a variant of the so called mass conserving
mixed stress (MCS) method that has its origin in \cite{ledererthesis, lederer, lederer2}. In
the following, we discuss the fundamental ideas and give a precise
definition of the method and the finite element spaces that we use.
The main idea is to rewrite the incompressible Navier-Stokes equations
into a formulation which includes the (pseudo) stress and the rotations. 
For this let $\omega(\underline u) =
\mathrm{skew}(\nabla \underline u)$ be the rotation rate tensor, then
we have 
\begin{align*}
	\nabla \underline u = \epsilon(\underline{u}) + {\omega}(\underline u).
\end{align*}
Introducing the auxillary variables $\doubleunderline{\gamma} =
{\omega}(\underline u)$ and $\doubleunderline{\sigma} =
2\nu\epsilon(\underline{u})$, we can rewrite the original system
\eqref{NSE} as 
\begin{subequations} \label{MCS_cont}
\begin{alignat}{2}
	-\frac{1}{2\nu}{\doubleunderline{\sigma}} + \nabla\underline{u} - \doubleunderline{\gamma} &= \doubleunderline{0} && \qquad \textnormal{in}\:\Omega\times[0,T_e],  \label{MCS_cont_S}\\
	\frac{\partial\underline{u}}{\partial t} - \nabla\cdot\doubleunderline{\sigma} + (\underline{u}\,\cdot\,\nabla)\underline{u} + \nabla p &= \underline{f} && \qquad \textnormal{in}\:\Omega\times[0,T_e], \label{MCS_cont_M} \\
\nabla\,\cdot\,\underline{u}&=0 && \qquad\textnormal{in}\:\Omega\times[0,T_e], \\	
\mathrm{skew}(\doubleunderline{\sigma})&=0 && \qquad \textnormal{in}\:\Omega\times[0,T_e], \\	
	\underline{u}&=\underline{u}_{0} && \qquad \textnormal{in}\:\Omega, t=0,  \\
	\underline{u}&=0 & & \qquad \textnormal{in}\:\partial\Omega_{W}\times[0,T_e].
\end{alignat}
\end{subequations}
Note that the constraint $\mathrm{skew}(\doubleunderline{\sigma})=0$
assures symmetry of $\doubleunderline{\sigma}$, and is used below to
enforce weak symmetry also of the approximate stress. This approach is
very common for mixed finite element methods for elasticity, see for
example \cite{FdV2, PEERS, family} and very recently \cite{ls2}. In
addition, it is worth noting that 
\begin{align} \label{sigmatrace}
	\sigma = \dev(\doubleunderline \sigma), \quad \textrm{since} \quad
\tr(\doubleunderline{\sigma}) = \nabla \cdot u = 0.
\end{align}

In \cite{ledererthesis} the author showed how the underlying Stokes system
in \eqref{MCS_cont} can be discretized by means of using an
$H(\mathrm{div})$-conforming approximation space for the velocity. We
proceed similarly and choose the finite element spaces for the
velocity and pressure by 
\begin{align} \label{disc_velspace}
 V_h= RT^k(\mathcal{T}_h) \cap H_{0}(\mathrm{div}, \Omega), \quad \textrm{and} \quad  Q_h= \mathbb{P}^{k}(\mathcal{T}_h) \subset L^2(\Omega).
\end{align}
Note that we have the compatibility (or kernel preserving, see
\cite{boffi2013mixed}) condition 
\begin{align} \label{divVeqQ}
	\nabla\,\cdot\,V_h=Q_h. 
\end{align}
It remains
to define spaces for the approximation of the stress/strain and the
rotations. To motivate our choice, we consider well posedness of a
variational formulation (on the continuous level) of
\eqref{MCS_cont_M} with test functions $\underline v \in
H_0(\mathrm{div}, \Omega)$. If $\nabla \cdot \doubleunderline \sigma$ is
an element of $L^2(\Omega, \mathbb{R}^3)$, the second term in
\eqref{MCS_cont_M} can be written as $(\nabla \cdot \doubleunderline
\sigma, \underline v)$. Unfortunately, this does not lead to a
stable formulation. Alternatively, one could introduce less regularity
of the divergence of the stress by demanding that $\nabla \cdot
\doubleunderline \sigma$ can just continuously act (in the sense of a
functional) on $\underline v$. Thus the second term would read as
$\langle \nabla \cdot \doubleunderline \sigma, v \rangle$, where
$\langle \cdot, \cdot \rangle$ denotes the duality pair. Indeed, this
allows to show stability (see \cite{ledererthesis}) where $\doubleunderline{
\sigma}$ is an element of
\begin{align*}
	H(\mathrm{curl}\,\mathrm{div},\Omega) = \{\doubleunderline{\tau}\in L^2(\Omega,\mathbb{R}^{3\times 3} ): \tr(\doubleunderline{\tau}) = 0, \nabla\,\cdot\,\doubleunderline{\tau} \in  H_0(\mathrm{div},\Omega)^*\}.
\end{align*}
Here $H_0(\mathrm{div},\Omega)^*$ denotes the dual space. The
trace-free condition $\tr(\doubleunderline{\tau}) = 0$ is motivated by \eqref{sigmatrace}. To
approximate functions in $H(\mathrm{curl}\,\mathrm{div},\Omega)$, it
was then shown that the discrete functions need to be
normal-tangential continuous. This motivates to define the finite
element space 
\begin{equation*}
	\{ \doubleunderline{\tau}_h \in \mathbb{P}^{k+1}(\mathcal{T}_h,\mathbb{R}^{3 \times 3}): \, \tr(\doubleunderline{\tau}_h) = 0,   (\doubleunderline{\tau}_h)_{nt} \in \mathbb{P}^k(E, \mathbb{R}^3),  \llbracket  (\doubleunderline{\tau}_h)_{nt} \rrbracket = 0 \,  \forall \; E\in\mathcal{F}_h^I\}.
\end{equation*}
This reads as the space of element-wise matrix-trace-free
matrix-valued polynomials of order $k+1$, whose normal-tangential
trace (on facets) is a vector-valued polynomial of order $k$ and whose
normal-tangential jump vanishes. Note that the additional
normal-tangential bubbles, i.e. those polynomials that are of order
$k+1$ but have a normal-tangential trace of order $k$, are only needed
for stability of the method which follows with the same lines as in
\cite{lederer, lederer2}. Instead of incorporating normal-tangential
continuity into the space as above, we use an additional hybrid
variable (on the skeleton $\mathcal{F}_h$), that will be used as a
Lagrange multiplier, see \cite{refId0}. This has the advantage that
the (local element-wise) degrees of freedom of the stress
approximation are decoupled, and thus can be locally eliminated (see
Section 4.2 in \cite{lederer3}). Overall this leads to the definition
of 
\begin{align*}
	\Sigma_h &=\{ \doubleunderline{\tau}_h \in \mathbb{P}^{k+1}(\mathcal{T}_h,\mathbb{R}^{3 \times 3}): \tr(\doubleunderline{\tau}_h) = 0,   (\doubleunderline{\tau}_h)_{nt} \in \mathbb{P}^k(E, \mathbb{R}^3)\; \forall E\in\mathcal{F}_h^I\},\\
	\hat{V}_h&=\{\hat{\underline{v}}_h\in \mathbb{P}^{k}(\mathcal{F}_h, \mathbb{R}^3):\hat{\underline{v}}_h\,\cdot\,\underline{n}=0 \; \forall E\in\mathcal{F}_h, \; \hat{\underline v} = 0 \; \forall E\in\mathcal{F}^D_h\}.
\end{align*}
It is important to note that by the above choice, the normal-tangential jump of functions from the "broken" stress space $\Sigma_h$
are elements of $\hat{V}_h$, i.e. 
\begin{align} \label{jumpinVhat}
	\llbracket  (\doubleunderline{\tau}_h)_{nt} \rrbracket \in \hat V_h \quad \forall \tau_h \in \Sigma_h, \forall E \in \mathcal{F}_h.
\end{align}
Additionally note that the symbol $\hat{V}_h$ for the space of the
Lagrange multipliers (used to incorporate the normal-tangential
continuity of the stress approximation) was chosen, since functions in
$\hat{V}_h$ correspond to approximations of the tangential trace of
the (exact) velocity solution, compare \cite{refId0}, or Section 7.2.2
in \cite{boffi2013mixed}.
It remains to define a space for the approximation of
$\doubleunderline{\gamma}$ which is given by 
\begin{equation*}
W_h=\mathbb{P}^{k}_{skew}(\mathcal{T}_h,\mathbb{R}^{3 \times 3}) =
\{ \doubleunderline{\eta} \in
\mathbb{P}^{k}(\mathcal{T}_h,\mathbb{R}^{3 \times 3}) :
(\doubleunderline{\eta} + \doubleunderline{\eta}^T)= 0 \: \forall
T\in\mathcal{T}_h \}. 
\end{equation*}

The (semi-)discrete hybrid MCS method with weakly imposed symmetry then reads as:
Find
$(\doubleunderline{\sigma}_h,\underline{u}_h,\underline{\hat{u}}_h,\doubleunderline{\gamma}_h,p_h)
\in \Sigma_h \times V_h \times \hat{V}_h \times W_h \times Q_h$ such
that,
\begin{subequations} \label{MCS}
\begin{align*}
a_h (\doubleunderline{\sigma}_h,\doubleunderline{\tau}_h) + b_{2h}(\doubleunderline{\tau}_h,(\underline{u}_h,\underline{\hat{u}}_h,\doubleunderline{\gamma}_h))&=0 & & \forall \doubleunderline{\tau}_h \in \Sigma_h, \numberthis \label{MCS_sigma} \\ 
(\frac{\partial \underline{u}_h}{\partial t},\underline{v}_h) + b_{2h}(\doubleunderline{\sigma}_h,(\underline{v}_h,\underline{\hat{v}}_h,\doubleunderline{\eta}_h)) \numberthis \label{MCS_u} & & & \\ 
+ b_{1h}(\underline{v}_h,p_h)+c_h(\underline{u}_h,\underline{u}_h,\underline{v}_h)&=(\underline{f},\underline{v}_h) & &\forall (\underline{v}_h,\underline{\hat{v}}_h, \doubleunderline{\eta}_h) \in V_{h} \times \hat{V}_{h} \times W_h, \\
b_{1h}(\underline{u}_h,q_h)&=0 \numberthis \label{MCS_p}& &\forall q_h\in Q_h.
\end{align*}
\end{subequations}
The bilinear forms are defined as follows. The symmetric form which  
includes the stress variables, and the incompressibility constraint
are given by
\begin{align*}
a_h (\doubleunderline{\sigma}_h,\doubleunderline{\tau}_h) &= \sum_{T\in\mathcal{T}_h} \int_T -\frac{1}{2\nu}\:\doubleunderline{\sigma}_h\,:\,\doubleunderline{\tau}_h\:d\underline{x},\\
b_{1h}(\underline{u}_h,q_h) &=-\sum_{T\in\mathcal{T}_h}\int_T(\nabla\,\cdot\,\underline{u}_h)q_h\:d\underline{x} = -(\nabla\,\cdot\,\underline{u}_h,q_h),
\end{align*}
respectively, where the last equation follows since $\underline{u}_h$
is normal continuous. Note that due to \eqref{divVeqQ}, equation
\eqref{MCS_p} enforces an exactly divergence-free property of the
discrete velocity solution $\underline u_h$. 
The other constraint is given by
\begin{align*}
b_{2h}(\doubleunderline{\sigma}_h,(\underline{v}_h,\underline{\hat{v}}_h,\doubleunderline{\eta}_h)) 
= \numberthis \label{b2h} & \sum_{T\in\mathcal{T}_h} 
\bigg( -\int_T (\nabla\,\cdot\,\doubleunderline{\sigma}_h)\,\cdot\,\underline{v}_h\:d\underline{x} 
+ \int_{\partial T}  (\doubleunderline{\sigma}_h)_{nn} (\underline{v}_h\,\cdot\,\underline{n})\:d\underline{s} 
\\ & - \int_T \doubleunderline{\sigma}_h\, : \, \doubleunderline{\eta}_h\:d\underline{x} 
+ \int_{\partial T}  (\doubleunderline{\sigma}_h)_{nt}\,\cdot\,\underline{\hat{v}}_h\:d\underline{s} \bigg).
\end{align*}
Whereas the first and the second term can be interpreted as a
discrete distributional divergence (see also the explanation and
equation (4.2) in \cite{lederer2}), i.e. a discrete version of the
term $\langle \nabla \cdot \doubleunderline{\sigma}, \underline v
\rangle$ as mentioned above, the third integral enforces weak
symmetry. The last integral enforces normal-tangential continuity
since from \eqref{MCS_u} we have
\begin{align*}
	\sum_{T\in\mathcal{T}_h} \int_{\partial T}  (\doubleunderline{\sigma}_h)_{nt}\,\cdot\,\underline{\hat{v}}_h\:d\underline{s}
	= \sum_{F \in \mathcal{F}_h} \int_F \llbracket  (\doubleunderline{\sigma}_h)_{nt} \rrbracket \,\cdot\,\underline{\hat{v}}_h\:d\underline{s} = 0 \quad \forall \underline{\hat{v}}_h \in \hat V_h,
\end{align*}
and thus by \eqref{jumpinVhat} we have $\llbracket  (\doubleunderline{\sigma}_h)_{nt} \rrbracket = 0$.

The non-linear convective term is the same as in \cite{lehrenfeld} and is 
given by 
\begin{align}
c_h(\underline{u}_h,\underline{u}_h,\underline{v}_h)= & \sum_{T\in\mathcal{T}_h}\int_T(\nabla\underline{u}_h\cdot\underline{u}_h)\cdot\underline{v}_h\:d\underline{x} \label{upwind} \\
 &+\sum_{E\in\mathcal{F}_h} \bigg(- \int_{E}(\underline{u}_h\,\cdot\,\underline{n})\llbracket\underline{u}_h\rrbracket\,\cdot\,\langle\underline{v}_h\rangle\:d\underline{s} \nonumber \\
&+\frac{1}{2}(\underline{u}_h\,\cdot\,\underline{n})\llbracket\underline{u}_h\rrbracket\,\cdot\,\llbracket\underline{v}_h\rrbracket\:d\underline{s}\bigg).\nonumber
\end{align}
The last integral results in a standard (DG) upwind stabilization, see for example \cite{MR2165335}.
If this term is left out, we have no convection stabilization, i.e. a central flux
is used.

In \eqref{MCS} the Dirichlet no-slip conditions \eqref{MCS_cont} were
split into tangential and normal parts. While $\underline u_h \cdot
n = 0$ is incorporated into the space $V_h$, see
\eqref{disc_velspace}, the vanishing tangential trace is incorporated
into the facet variables in $\hat V_h$. A detailed discussion on the
boundary conditions (including Neumann and mixed conditions) is given
in Section 4 in \cite{ledererthesis}.

To derive the fully discrete system, we use the second-order
Runge-Kutta scheme ARS(2,2,2) of the implicit-explicit (IMEX) method
introduced in \cite{IMEX} as it was also used in \cite{lehrenfeld}.
Therein the stiff parts of the system, i.e. the diffusive terms that
include the bilinear-forms $a_h(\cdot,\cdot)$ and $b_{2h}(\cdot,
\cdot)$, and the terms related to the incompressibility constraint
$b_{1h}(\cdot, \cdot)$, are treated implicitly, and the convective
term is always treated explicitly. Note that this is essential in
order to maintain the exact divergence-free property of the discrete
velocity solution. A crucial advantage of this approach is that no
non-linear solver is needed. Instead, we only need to assemble and
factorize the matrix that corresponds to the linear parts of
\eqref{MCS} once, which results in a very efficient
implementation. As
already discussed above, this elimination is only possible due to the
hybridization of the system by means of the space $\hat V_h$. 

Finally, note that numerical quadrature is applied to all bilinear
forms such that exact integration is performed. This is of crucial
importance for the stability of the method in turbulent flow regime.
We use $k$ Gaussian quadrature points in each direction for
$a_h(\cdot,\cdot)$, $k+1$ points are sufficient for
$b_{1h}(\cdot,\cdot)$ and $b_{2h}(\cdot,\cdot)$ and $\frac{3(k+1)}{2}$
points are required for the convective term $c_h(\cdot,\cdot,\cdot)$.






\section{Methodology}\label{methods}

In the following, we introduce the three different methodologies that
we consider in this work. First, we discuss the extension of classical
VMS methods in combination with the MCS discretization introduced in
the previous section. Then, we consider the ILES approach (i.e. no
explicit turbulence model is used) and finally we introduce a novel
idea which can be understood as a VMS method applied to the
intrinsic stabilization mechanisms of the ILES setting.

\subsection{Variational multiscale}\label{vms}

Since its introduction in \cite{VMShughes}, the variational multiscale
finite element method has been applied to several problems appearing
in computational fluid dynamics. This framework has been originally
developed to stabilize methods in problems where stability is not
ensured for high Reynolds number flows. The origin can be found in
the introduction of LES, which consists of resolving the large scale
structures and model the non-resolving part of the turbulent flow.
Occasionally, by using the filtered Navier-Stokes equations, the new
appearing subgrid stress tensor term leads to a closure problem. This
mathematical term represents the effect of the missing unresolved
scales on the resolved ones. The Boussinesq assumption allows to solve
the closure problem and introduces the most common type of models, the
eddy viscosity models. The modeled subgrid stress tensor acts purely
as an additional dissipation mechanism in the system.

The classic explicit LES framework has drawbacks such as commutation
error and restricting theory of LES in unbounded domains
\cite{sagaut}. One remarkable disadvantage is that some explicit
turbulence models may produce unphysical dissipation in the laminar
regime, thus without existence of subgrid scales. Further on, the
modeled impact of unresolved scales is incorporated into the entire
range of resolved scales, whereas adding the effect on the largest
scale structures may lead to undesired behaviour of the flow.
Therefore, the VMS proposes an alternative to the standard LES
turbulence modeling technique. It allows to separate the range of flow
scales and hence the numerical scheme can be utilized for different
scale subranges. The main idea in VMS is to only apply the effect of
the non-resolving structures to the small resolved scales, therefore
allow the large resolved eddies to be untouched. Nevertheless, the
large range of scales are still influenced indirectly by the model
because of the inherent coupling of all scales. Many shortcomings
(such as commutation error) of the traditional LES are avoided by the
VMS approach. LES results often show overexceeding stabilization
behaviour, where the VMS allows more fine-grained stabilization
effects at the higher frequent modes.

Several versions of VMS-type implementations have been proposed in the
literature. In this paper, the three-scale projection-based VMS method
from \cite{VMS,Volker} is considered and is explained in the
following.

To explain the basic concept, we consider a standard $H^1$-conforming
discrete velocity space $V_h$ and will then explain how the ideas can
be extended to the MCS method from the previous section. The idea 
is to split the velocity space into (fictitious) spaces used to
approximate large, small resolved and unresolved scales respectively,
i.e. we consider
\begin{equation*}
V_h=V^L_h\oplus V^S_h\oplus V^U_h.
\end{equation*}
Applying this decomposition, we can then write the solution and test functions as
\begin{align*}
\underline{u}_h=\underline{u}^L_h+\underline{u}^S_h+\underline{u}^U_h,\quad \textrm{and} \quad
\underline{v}_h=\underline{v}^L_h+\underline{v}^S_h+\underline{v}^U_h.
\end{align*}
Following \cite{Volker}, one motivates an additional viscosity, called
the eddy viscosity, which appears in the diffusive part related to
$\underline{u}^S_h$ in the momentum equation. More precisely, a finite
element approximation (based on the space $V_h$ from above) of the
momentum equation \eqref{NSE_M} then includes the additional term 
\begin{equation} \label{eq::additionalvisc}
    (2\nu_t\epsilon(\underline{u}^S_h),\epsilon(\underline{v}^S_h)),
    \end{equation}
which  models the influence of the unresolved scales onto the small
resolved ones and ignoring direct effect of the large resolved
structures. Of course, additional modeling is needed to derive a
proper definition of $\nu_t$, commonly done via an SGS model.
Since a direct decomposition of $V_h$ is not desired, a crucial
question when considering a VMS method is how to incorporate
\eqref{eq::additionalvisc}.
In this work, we use a coarse space projection-based method where
\eqref{eq::additionalvisc} is written as
\begin{equation} \label{eq::additionalviscimpl}
\big(2\nu_t\epsilon(\underline{u}^S_h),\epsilon(\underline{v}^S_h))\big) \cong \big(2\nu_t\epsilon(\underline{u}_h),\epsilon(\underline{v}_h)\big)-\big(2\nu_t\Pi_{L_h}\epsilon(\underline{u}_h),\epsilon(\underline{v}_h)\big),
\end{equation}
with some projection operator $\Pi_{L_h}$. Note, that the right hand
side only includes the velocity $\underline{u}_h$ and not
$\underline{u}^S_h$. A common choice of $\Pi_{L_h}$ is the
$L^2$-projection into the space $L_h$ given by
\begin{equation*}
    L_h = \mathbb{P}^{m}(\mathcal{T}_h,\mathbb{R}^{3\times 3}) \cap L \quad \textrm{with} \quad L = \{\doubleunderline{l}\in L^2(\Omega,\mathbb{R}^{3\times 3}):\doubleunderline{l}=\doubleunderline{l}^T\},
\end{equation*}
which reads as element-wise symmetric matrix-valued polynomials of a
given fixed order $m$. The VMS method is less sensitive to the choice
of the SGS model (used to define $\nu_t$) than in traditional LES.
This is due to the fact that the model influences only a subrange of
scales which can be controlled by the order $m$ of the space $L_h$. On
the one hand, if $L_h=\{0\}$ (i.e. the last term in
\eqref{eq::additionalviscimpl} vanishes), the classic LES model is
recovered. On the other hand, if $m$ is chosen high enough such that
$\{\epsilon(\underline{v}_h):\underline{v}_h\in V^h\} \subset L_h$, the
modeled viscosity is switched off (the right hand side in
\eqref{eq::additionalviscimpl} is zero) and the no-model approach is
reobtained. Summarizing, a small order $m$ results in a bigger impact
of the turbulence model, while a higher order $m$ results in no model
at all.



In the following, we discuss how the VMS approach can be extended to
the MCS method from the previous chapter. For this purpose, let $V_h$ be again
the normal continuous Raviart-Thomas space, see \eqref{disc_velspace}.
The main idea is to apply the projection based splitting from
\eqref{eq::additionalviscimpl} already in the original definition of
the auxillary variable $\doubleunderline{\sigma}$ in
\eqref{MCS_cont_S}. We define
\begin{equation*}
    \doubleunderline{\sigma} = 2\big( (\nu+\nu_t)\epsilon(\underline{u}) - \nu_t\doubleunderline{g} \big) ,
    \end{equation*}
where $g$ is some additional strain tensor.
Reformulating above equation gives 
\begin{equation}
\epsilon(\underline{u}) = \frac{1}{2(\nu+\nu_t)}\big( \doubleunderline{\sigma} + 2\nu_t\doubleunderline{g} \big), \label{epsilon}
\end{equation}
thus, testing with a test function $\doubleunderline{\tau}$ and integrating over the domain we have
\begin{align} \label{eq::contVMSMCS_one}
    \int_\Omega \frac{1}{2(\nu+\nu_t)} ( \doubleunderline{\sigma} + 2\nu_t\doubleunderline{g} ):\doubleunderline{\tau}\:d\underline{x} - \int_\Omega \epsilon(\underline{u}):\doubleunderline{\tau}\:d\underline{x} = 0.
\end{align}
In addition (motivated by the findings from
\eqref{eq::additionalviscimpl}), we define $g$ as the $L^2$ projection
in $L$ by
\begin{align} \label{eq::contVMSMCS_two}
    \int_\Omega (\epsilon(\underline{u}) - \doubleunderline{g}) : \doubleunderline{l} \:d\underline{x} = 0 \quad \forall \doubleunderline{l} \in L.
\end{align}
Note that these equations are still formulated on the continuous
level. To derive the discrete method, we continue as follows. In a
first step, we insert \eqref{epsilon} into \eqref{eq::contVMSMCS_two},
and substitute the continuous functions by their discrete counterpart
in $\Sigma_h$ and $L_h$, which results in 
\begin{equation*}
    \sum_{T\in\mathcal{T}} \bigg( \int_T \frac{1}{2(\nu+\nu_t)} \big( \doubleunderline{\sigma}_h+2\nu_t \doubleunderline{g}_h \big):(2\nu_t\doubleunderline{l}_h)\:d\underline{x} - \int_T \doubleunderline{g}_h:(2\nu_t\doubleunderline{l}_h)\:d\underline{x} \bigg) = 0.
    \end{equation*}
Thus, adding the discrete version of \eqref{eq::contVMSMCS_one}, again
interpreting the last integral in a (discrete) distributional way, see
discussion below \eqref{b2h}, we have with the bilinear form 
\begin{align*}
a_{h}^{VMS}((\doubleunderline{\sigma}_h,\doubleunderline{g}_h),(\doubleunderline{\tau}_h,\doubleunderline{l}_h))=& \sum_{T\in\mathcal{T}} \bigg( \int_T \frac{1}{2(\nu+\nu_t)} \big( \doubleunderline{\sigma}_h+2\nu_t \doubleunderline{g}_h \big):(\doubleunderline{\tau}_h+2\nu_t\doubleunderline{l}_h)\:d\underline{x} \\
&-\int_T \doubleunderline{g}_h:(2\nu_t\doubleunderline{l}_h)\:d\underline{x}\bigg).
\end{align*}
the following MCS VMS formulation:\\
Find
$(\doubleunderline{\sigma}_h,\doubleunderline{g}_h,\underline{u}_h,\underline{\hat{u}}_h,\doubleunderline{\gamma}_h,p_h)
\in \Sigma_h \times L_h \times V_h \times \hat{V}_h \times W_h \times
Q_h$ such that,
\begin{align*}
a_{h}^{VMS}((\doubleunderline{\sigma}_h,\doubleunderline{g}_h),(\doubleunderline{\tau}_h,\doubleunderline{l}_h)) & & & \\
+ b_{2h}(\doubleunderline{\tau}_h,(\underline{u}_h,\underline{\hat{u}}_h,\doubleunderline{\gamma}_h))&=0 & & \forall (\doubleunderline{\tau}_h, \doubleunderline{l}_h) \in \Sigma_h \times L_h, & \\
(\frac{\partial \underline{u}_h}{\partial t},\underline{v}_h) + b_{2h}(\doubleunderline{\sigma}_h,(\underline{v}_h,\underline{\hat{v}}_h,\doubleunderline{\eta}_h)) & & & \\
+ b_{1h}(\underline{v}_h,p_h)+c_h(\underline{u}_h,\underline{u}_h,\underline{v}_h)&=(\underline{f},\underline{v}_h) & &\forall (\underline{v}_h,\underline{\hat{v}}_h, \doubleunderline{\gamma}_h) \in V_{h} \times \hat{V}_{h} \times W_h,& \\ 
b_{1h}(\underline{u}_h,q_h)&=0 & &\forall q_h\in Q_h.&
\end{align*}

For this study, we used the WALE turbulence model from \cite{WALE} for
calculating the turbulent viscosity $\nu_t$. The WALE model is based
on the square of the velocity gradient tensor, which takes into
account the stress tensor as well as the rotation tensor. It takes
near-wall turbulence into consideration, without the use of any
dynamic models or damping functions. To this end, we define
\begin{equation*}
\nu_t = (C_w\Delta)^2\frac{S^2(\underline{u}_h)^{\frac{3}{2}}}{\epsilon^2(\underline{u}_h)^{\frac{5}{2}} + S^2(\underline{u}_h)^{\frac{5}{4}}},
\end{equation*}
with 
    $
\epsilon^2(\underline{u}_h) = \epsilon(\underline{u}_h):\epsilon(\underline{u}_h),
$
and
\begin{align*}
S^2(\underline{u}_h) = &\frac{1}{6}\big(\epsilon^2(\underline{u}_h)\epsilon^2(\underline{u}_h) + \omega^2(\underline{u}_h)\omega^2(\underline{u}_h)\big)+\frac{2}{3}\epsilon^2(\underline{u}_h)\omega^2(\underline{u}_h) \\ &+2\tr\big(\epsilon(\underline{u}_h)\epsilon(\underline{u}_h)\omega(\underline{u}_h)\omega(\underline{u}_h)\big).
\end{align*}
Here, $C_w = 0.5$ is a model constant and the length scale is defined as $\Delta = \frac{\sqrt[3]{\Delta x \Delta y \Delta z}}{k+1}$.

Regarding to the numerical quadrature of the MCS VMS method, the
polynomial order has to be increased for $a_{h}^{VMS}(\cdot,\cdot)$ in
order to obtain exact integration. Since the update matrix does not
keep constant coefficients anymore ($\nu_t$ varies over time), we
recompute the corresponding system matrix after a given (fixed) time
interval. This highly reduces the computational effort and has shown
to be of negligible impact on the simulation. For simplicity reasons,
the polynomial order of the $L_h$ space, which defines the scale
separation, is set to be equal for every element.
 
\subsection{ILES} 

As mentioned in section \ref{intro}, the general question of
turbulence modeling is how to account for the unresolved subgrid
scales in the best way. Conventional LES methods have shown that
additional dissipation ensures stability and appropriate results. By
introducing more sophisticated SGS models, many insufficiencies of LES
could be avoided and increased the accuracy of computations of more
complex flows. The VMS does not show much dependence on the chosen
turbulence model, as the intrinsic scale separation overcomes many of
the drawbacks of classical LES and reduces the total amount of
additional model dissipation in the system.

The very flexible and stable characteristic of DG discretization
methods has made the stabilization requirement, i.e. the turbulence
model used in LES/VMS obsolete. By this, it makes the ILES approach
attractive since the theoretical and computational complexities of
classical turbulence modeling are avoided. A DG formulation often
enforces stability by using upwind Riemann fluxes which introduces
numerical dissipation by means of additional jump terms, as can be
seen in the definition of the convective trilinear form
\eqref{upwind}. It is recognized that this is the main source for
providing stability in under-resolved flow regimes. The idea here is
that in contrast to laminar flows, the jump terms in an under-resolved
turbulent flow state get significantly bigger which introduces an
additional (local) diffusion. A linear dispersion-diffusion analysis
in \cite{moura2} justified the capabilites of general DG methods for
under-resolved turbulent flow simulation.

For MCS ILES, the standard formulation of section \eqref{mcs} is taken. 
 
\subsection{High-order projection-based upwind ILES}\label{hopu_chapter}

For highly under-resolved simulations, we obtain large amount of
numerical dissipation in the near-wall region due to the turbulent
boundary layer. This excessive dissipation is not necessarily needed
for a stable solution and origins from the upwind term from equation
\eqref{upwind}. This term acts as a convection stabilization in order
to stabilize high fluctuating turbulent flows. In principle, upwinding
is an anisotropic mechanism, however, the turbulent mixing properties
of the flow lead to the fact that it acts more isotropically. If any
subgrid scales appear in the flow, the effect of those non-resolved
structures is taken into account by the velocity jump at the element
interfaces. Note, that we use an $H(\mathrm{div})$-conforming method,
therefore only the tangential component of the velocity produces jumps
at element interfaces. 

As mentioned above, the last integral (resulting from the chosen
upwind Riemann solver) in \eqref{upwind} produces artificial
dissipation that is proportional to the velocity jumps. A numerical
investigation shows that the magnitude of these jumps are in general
larger in the near-wall region than in the outer area of the flow.
This effect results into excessive amount of additional dissipation in
the proximity of walls. To overcome this problem, we introduce  the so
called high-order projection-based upwind method (HOPU) for ILES,
which follows a similar idea as used in the VMS approach. The upwind
term penalizes the whole range of resolved scales in the flow regime.
Typically, areas with high turbulent kinetic energy lead to an
increased jump size at the element interfaces. In order to avoid
influencing large resolved structures in such areas, we introduce a
projection-based scale separation of the jumps. Then, the numerical
dissipation only acts on the high-order parts of the flow field. This
is done by the introduction of an additional $L^2$-projection in the
last integral of \eqref{upwind}. Thus, the low-frequent part of the
velocity (jump) is not taken into consideration. Consequently,
low-order terms are considered in a central flux manner.

To this end, we define an additional space 
\begin{equation*}
\hat{V}_h^{proj}=\{\underline{v}_h\in \mathbb{P}^l(\mathcal{F}_h, \mathbb{R}^3):\underline{v}_h\,\cdot\,\underline{n}=0 \; \text{on} \; E\in\mathcal{F}_h\},
\end{equation*} 
where the polynomial order is chosen such that  $0 \leq l \leq k$.
Then, the convective term is changed to 
\begin{align*}
    c^{HOPU}_h(\underline{u}_h,\underline{u}_h,\underline{v}_h)=  &\sum_{T\in\mathcal{T}_h}\int_T(\nabla\underline{u}_h\cdot\underline{u}_h)\cdot\underline{v}_h\:d\underline{x}  \\
     &+\sum_{E\in\mathcal{F}_h} \bigg(- \int_{E}(\underline{u}_h\,\cdot\,\underline{n})\llbracket\underline{u}_h\rrbracket\,\cdot\,\langle\underline{v}_h\rangle\:d\underline{s} \\
  &+\int_{E}\frac{1}{2}(\underline{u}_h\,\cdot\,\underline{n})(I-\Pi^l_{\hat{V}})\llbracket\underline{u}_h\rrbracket\,\cdot\,\llbracket\underline{v}_h\rrbracket\:d\underline{s}\Big),
\end{align*}
where $\Pi^l_{\hat{V}}:L^2(E) \rightarrow \hat{V}_h^{proj}(E)$ is the
facet wise $L^2$-projection into $\hat{V}_h^{proj}(E)$. 

If $l=0$, we nearly reobtain the standard upwinding approach. In
practice, the lowest order part is of no relevance and therefore
negligible. On the other hand, if $l=k$, no upwinding takes place and
we obtain a central treatment of the convection. In principle, regions
with large velocity jumps should have a higher order upwind
projection. Areas with smaller jumps, are not affected a lot by the
convection stabilization and therefore a small order projection is
sufficient as it converges to the standard ILES method. Note, as the
resolution increases and a larger range of scales is covered, the
impact of the upwind mechanism is getting reduced since there holds
$\lim\limits_{h \rightarrow
0}{\llbracket\underline{u}_h\rrbracket}=0$.


A simple way to address the problem of determining the local
polynomial order used in $\Pi^l_{\hat{V}}$ is to divide the domain
into two regions. Thus, we treat the near-wall region with HOPU while
in the outer region we consider the standard upwind technique. This
approach works well for simple geometries i.e. channel flow problem,
but is rather difficult to predict in more complicated domains and in
the presence of complex flow phenomenons. Another difficulty, which is
not fulfilled by a simple wall layer, is that regions with high
turbulent kinetic energy may also occur in the outer flow field.

A more intrinsic way to solve this problem is to use an adaptive local
order version. For this purpose, we define a parameter to give an
estimate about the size of the jumps across element-interfaces. The
high order average absolute value jump is given by
\begin{equation*}
\eta = \frac{\int_{E} |(I-\Pi^l_{\hat{V}})\llbracket\underline{u}_h\rrbracket |\:d\underline{s}}{\int_{E} | \langle\underline{u}_h\rangle |\:d\underline{s}}.
\end{equation*}
Based on this local (spatial) average $\eta$, we can then decide which
polynomial order is used. We choose values $\eta_i$ such that
$\eta_0=0 < \eta_1 < \eta_2 < ... < \eta_{k+1}=1$. Then for $\eta_i <
\eta_{loc} \le \eta_{i+1}$ we use the local order $l_{loc}=i$. In our
computations the local polynomial order $l_{loc}$ was determined after
each time step and  changed appropriately. The adaptive HOPU variant will be
motivated in more detail in section \ref{channel}.

\section{Turbulent channel flow}\label{channel}

The 3D turbulent channel flow problem is a common test case problem
for assessing the ability of structure resolving flow solvers to deal
with the wall-bounded turbulence phenomena. We consider the
rectangular cuboid sized $\Omega = (0,2\pi) \times (0,2) \times
(0,\pi)$ for all flow simulations. In $x$-direction as well as
$z$-direction we use periodic boundary conditions. For $y=0$ and $y=2$
we consider no-slip boundary conditions in order to intimidate rigid
walls. The time step size $\Delta t = 0.001$ is used. The grid consists of $N$ hexahedral elements in each direction,
leading to a total number of $N^3$ elements per domain. Additionally,
as commonly done \cite{ILES_christoph}, the mesh is
stretched in $y$-direction to be able to resolve the sharp velocity
gradient in the boundary layer by means of the stretching function
\begin{equation*}
	\Phi(y) = \frac{\mathrm{tanh}(1.8(y-1))}{\mathrm{tanh}(1.8)}+1.
\end{equation*}
The flow characteristic is specified by the friction Reynolds number
$Re_t = u_tL_y/\nu = 392.24$ for all channel flow computations, where
$L_y=1$ is half the channel height and
$u_t=\sqrt{\frac{\tau_w}{\rho}}$ the friction velocity with the 
the wall shear stress  $\tau_w$ and  the density  $\rho=1$ of the fluid.

Usually, the fluid motion is driven by a bulk force $\underline{f} =
(f_x,\:0,\:0)^T$, whereas $f_x$ is dynamically adjusted to match the
aimed friction Reynolds number flow. However, this is not necessary
for MCS in general since it naturally preserves the wall stress for
every time step. It can be simply shown (by choosing the test function
$v_h = (1,0,0)^T \in V_h$) that 
\begin{align*}
\sum_{T\in\mathcal{T}}\int_T f_x \:d\underline{x} 
= \sum_{T\in\mathcal{T}}\int_{\partial T\cap\partial\Omega_{D}} (\doubleunderline{\sigma}_h)_{nt,x} \:d\underline{s},
\end{align*}
where $(\doubleunderline{\sigma}_h)_{nt,x}$ is the $x$-component of
the vector $(\doubleunderline{\sigma}_h)_{nt}$. Note that due to
\eqref{MCS_cont_S} we have that $(\doubleunderline{\sigma}_h)_{nt,x} =
\tau_w$, thus the MCS method naturally preserves the wall stress.
Therefore, in order to obtain an average friction Reynolds number over
a time interval, we do not have to use any bulk velocity or wall
stress control mechanism for solving the problem but simply set $f_x$
to the corresponding value.

The relevant quantities of interest are the three velocity components
$\underline{u}_h = (u, v, w)^T$, where we skipped the $h$-index for
readability. After a statistical-stationary state of the flow is
reached, sampling data of the velocity solution are taken. Our main
interest are the normalized mean of the velocity
$\underline{\overline{u}}^+ = \overline{\underline{u}_h}/u_t$ and the
normalized Reynolds stresses
$\overline{\underline{u}'_i\underline{u}'_j}^+=(\overline{\underline{u}_i\underline{u}_j}
- \overline{\underline{u}}_i\overline{\underline{u}}_j)/u_t^2$ where
$\underline{u}_i$ with $i=1,\ldots,3$ are the velocity components of
$\underline{u}_h$. The mean operator $\overline{(*)}$ involves
averaging over time and in the homogenous spatial directions. By doing
so, both quantities can be presented as functions over $y$.
Furthermore, wall units $y^+=yu_t/\nu$ are used. The energy spectrum
$E^+$ is obtained by discrete Fourier transformation of
$\underline{u}_h$ at planes orthogonal to $y$, averaged in time and
normalized by $u_t^2$.

\begin{table}
	\centering
	\begin{tabular}{|c|c|c|c|c|c|c|}
		\hline
		$N$/$k$ & 6/3 & 12/3 & 16/3 & 4/6 & 6/6 & 8/6\\ \hline
		gDOF$\cdot 10^{-3}$ & 43.2 & 331.0 & 775.7 & 63.6 & 208.1 & 485.6 \\
		$\Delta y^+_W$ & 11.8 & 4.3 & 3.0 & 13.6 & 6.7 & 4.3\\
		\hline
	\end{tabular}
	\vspace{3mm}
	\caption{Simulation parameters for the turbulent channel flow.}
	\label{table_channel}
\end{table}
 
\subsection{Comparison of results}\label{channel-compare}

We investigate the accuracy of the different methods (ILES, VMS, HOPU)
by comparing statistical and spectral quantities of the turbulent
channel flow to DNS reference data from \cite{DNSmoser}. Two different
polynomial degrees are chosen. The parameters of our  simulations are
shown in table \ref{table_channel}. Here, $\Delta y^+_W$ represents
the resolution of the nearest wall element. The globally coupled
degrees of freedom (gDOF) are defined as the number of DOFs that
remain in the system after an elimination (static condensation) of the
local DOFs that are associated to the variables
$\doubleunderline{\sigma}_h$ and $\doubleunderline{\gamma}_h$. We want
to mention that we did not take into consideration the additional
computational costs of the VMS (since the matrix is factorized more
than once due to the non-constant eddy viscosity) but compare each
approach on the same mesh with the same approximation order. Note that
although this favours the VMS in terms of accuracy vs. efficiency, no
significant improvement can be observed in the following comparison.

\begin{figure}[h]
	\centering
    \includegraphics[width=1\textwidth]{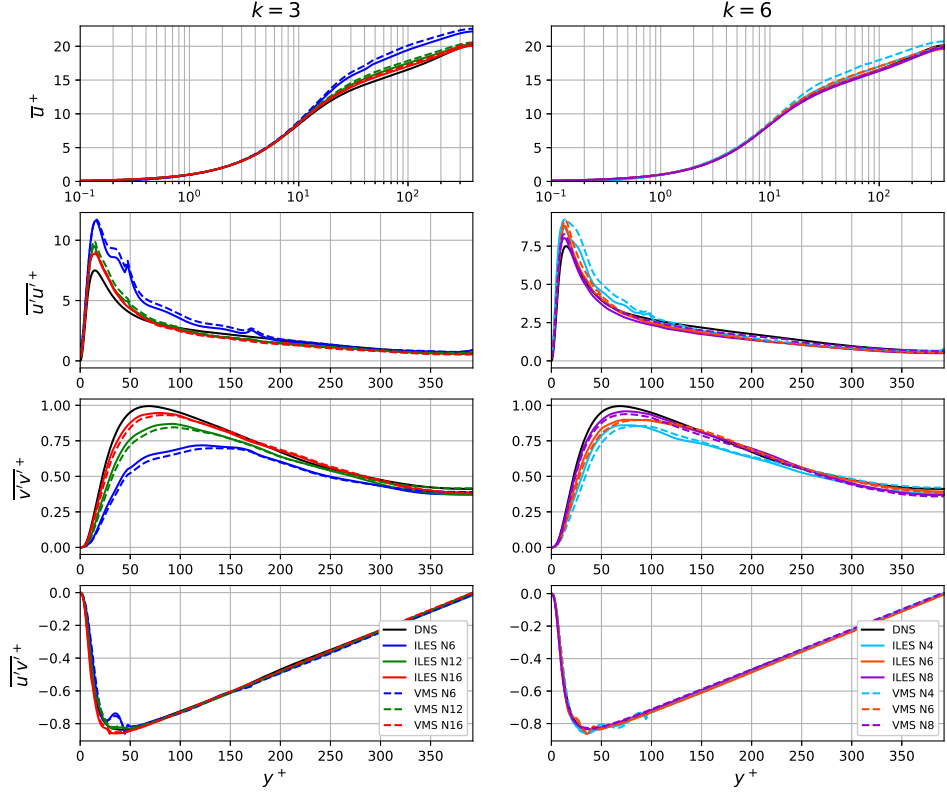}
    \caption{Normalized mean velocity $\overline{u}^+$ and normalized Reynolds stresses $\overline{u'u'}^+$, $\overline{v'v'}^+$ and $\overline{u'v'}^+$ over $y^+$.}
    \label{fig:vms}
\end{figure}

\begin{figure}[h]
	\centering
    \includegraphics[width=1\textwidth]{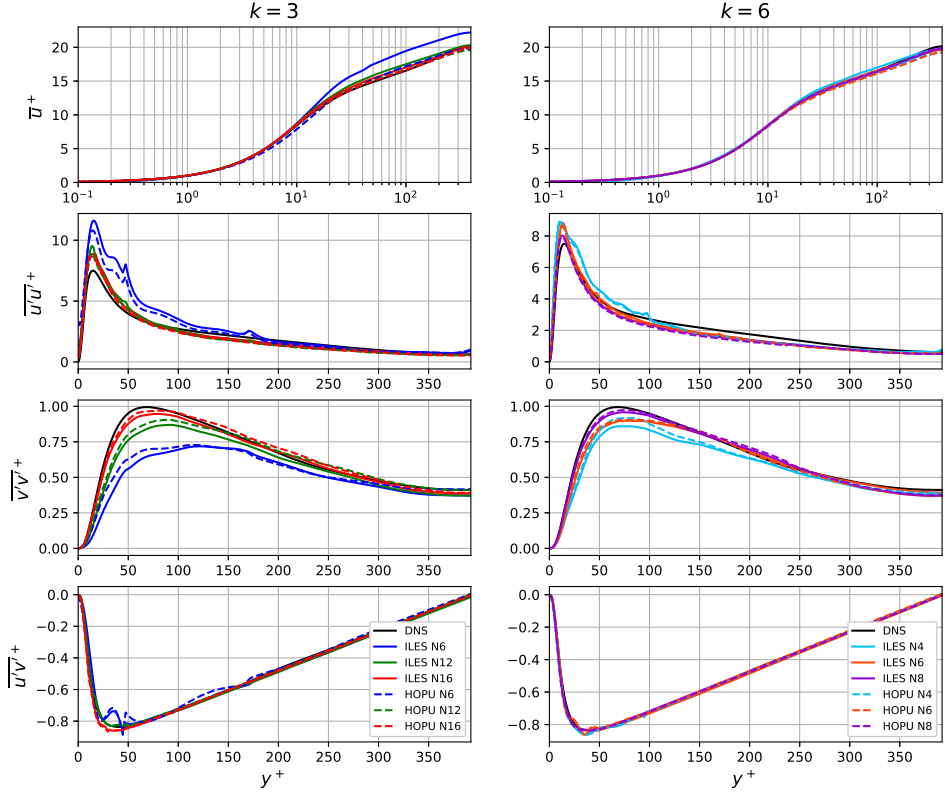}
    \caption{Normalized mean velocity $\overline{u}^+$ and normalized Reynolds stresses $\overline{u'u'}^+$, $\overline{v'v'}^+$ and $\overline{u'v'}^+$ over $y^+$.}
    \label{fig:hopu}
\end{figure}

Generally we observe for all test cases, see figure \ref{fig:vms},
figure \ref{fig:hopu} and table \ref{table_channel_results}, that by
increasing the spatial resolution, the results converge to the
reference data. Remarkably, even the highly under-resolved settings
deliver meaningful prediction of the involved quantities. The higher
order cases $k=6$ show overall better results than the lower order
ones. This is mainly due to two reasons. Firstly, if a low  and high
order method with roughly the same number of DOF is considered (thus
using a finer spatial resolution for the low order method), the higher
order simulation is able to capture a broader range of fine scales,
i.e. frequencies. Secondly, the inbuilt high-order dissipation
mechanism of the DG method allows dissipation for the high-frequent
modes only. Despite the roughly same number of DOF, the $k=3$ case
adds generally more numerical dissipation at larger scales to the
scheme than the $k=6$ case. In the following we discuss the results in
more detail. 

\textbf{Comparison of VMS and ILES}: For the VMS calculations we use
the polynomial orders $m=1$ (for $k=3$) and $m=3$ (for $k=6$) for the
large scale strain rate tensor space $L_h$. By comparing the ILES and
VMS simulations in figure~\ref{fig:vms}, we observe roughly the same
resulting data for the $N=\{12,16\}$/$k=3$ and $N=\{6,8\}$/$k=6$
cases. This is mainly due to the reason that the eddy viscosity
vanishes for increasing resolutions and by that the VMS approach
converges to the ILES.

The viscous sublayer ($y^+ < 5$) for $\overline{u}^+$ is accurately
approximated by all test cases. No eddy viscosity is applied by the
model in the laminar subregion. In the logarithmic layer and outer
region ($y^+ > 40$), the results show some differences for the highly
under-resolved case $N=6$/$k=3$ and $N=4$/$k=6$. For both setups, the
VMS simulations give slightly worse results for all quantities.
Additionally, the latter approach reduces the resolved turbulent
transport and in general damps more than the ILES. Therefore, the
buffer layer ($10 < y^+ < 40$) is not well resolved and results in a
small offset of $\overline{u}^+$.

\begin{figure}[h]
    \centering
    \includegraphics[width=0.7\textwidth]{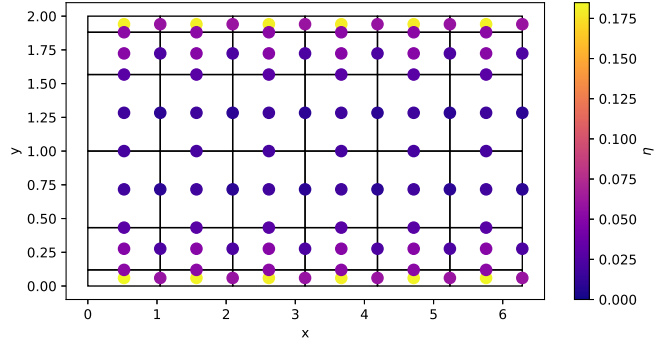}
    \caption{Average absolute value jump $\eta$ for the channel flow problem.}
    \label{fig:eta}
\end{figure}

\begin{figure}[h]
	\centering
    \includegraphics[width=0.7\textwidth]{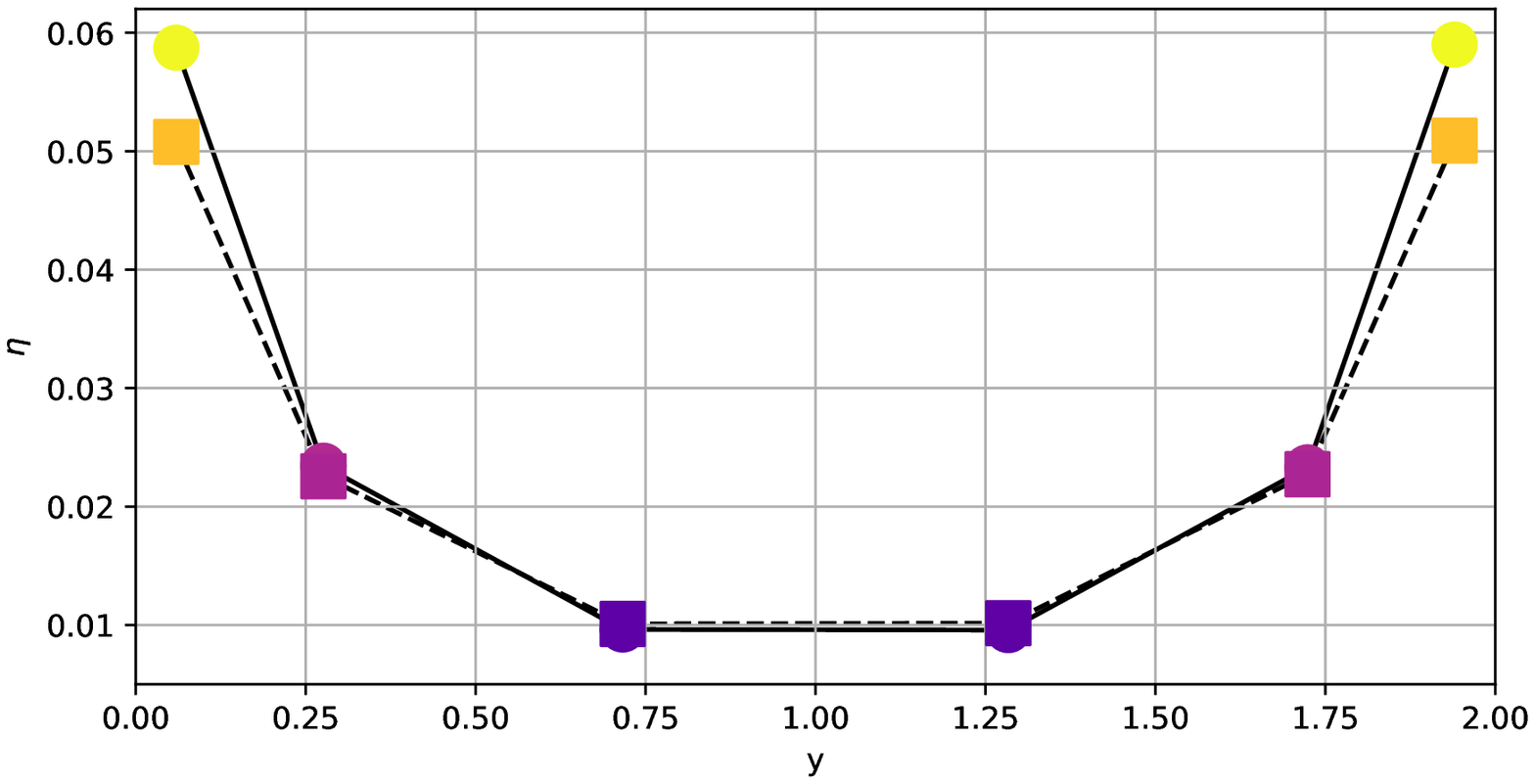}
    \caption{Average absolute value jump $\eta$ at $yz$-facets plotted over $y$ for the ILES (circles) and the HOPU variant (squares).}
    \label{fig:eta_curve}
\end{figure}

\textbf{Comparison of HOPU and ILES}: As mentioned at the end of
section~\ref{hopu_chapter}, one could choose either the standard or
the adaptive HOPU method. To motivate the usage of the latter, we plot
in figure \ref{fig:eta} the absolute value of the average value jump
$\eta$ for the channel flow problem using an ILES approach with
resolution $N=6/k=3$. 
The parameter $\eta$ is averaged over time and in $z$-direction and is
visualized in the midpoint of all facets. One observes that $\eta$ is
drastically increased close to the lower and upper wall. This
motivates to use an adaptive HOPU version with a high-order projection
closer to the walls and a low-order projection in the inner region. In
figure \ref{fig:eta_curve} we again plot $\eta$ but from a different
perspective where we also applied an average in $x$-direction. We
present the values for an adaptive HOPU approach as motivated above
(the exact choice of the local order limits is given below) with the
same resolution $N =6/k=3$ (squares) and the ILES from before
(circles).
We want to emphasize that $\eta$ represents only the high-order part
of the averaged jump when using HOPU ($l \ge 0$), but the full average
jump for the standard ILES ($\Pi^l_{\hat{V}}$ is removed). As we can
see, both approximations have roughly the same values in the inner
region of the channel. Since in that part of the channel the adaptive
HOPU method uses a low-order projection, this is the expected result.
Similarly, since a high-order projection is used closer
to the wall, we observe a bigger difference of $\eta$ there. This
shows, that the amount of additional (numerical) dissipation due to
the convection stabilization applied to the highest order functions,
is roughly the same in the inner part of the channel, but is reduced
closer to the wall when using the HOPU method. 






For the comparison of the velocity profiles and the Reynolds stresses
given in figure \ref{fig:hopu}, we have chosen an adaptive HOPU
approach with the limits $\eta = \{0.025,0.05,0.1\}$ for $k=3$ and
$\eta = \{0.005,0.015,0.03,0.07,0.12,0.2\}$ for $k=6$. 
A significant improvement of the mean velocity
$\overline{u}^+$ for the $k=3$ case, especially for $N=6$, is observed
for the new method. The HOPU version resolves the buffer region and
therefore gives a better approximation of the velocity field in the
logarithmic and outer area. The $k=6$ test cases do not show any
remarkable differences in the mean velocity and all resolutions give a
good prediction of the DNS data. This is because the jumps vanish for
highly resolved approximations, and thus naturally removes the
additional turbulence model due to HOPU. 

\begin{figure}[h]
	\centering
    \includegraphics[width=1\textwidth]{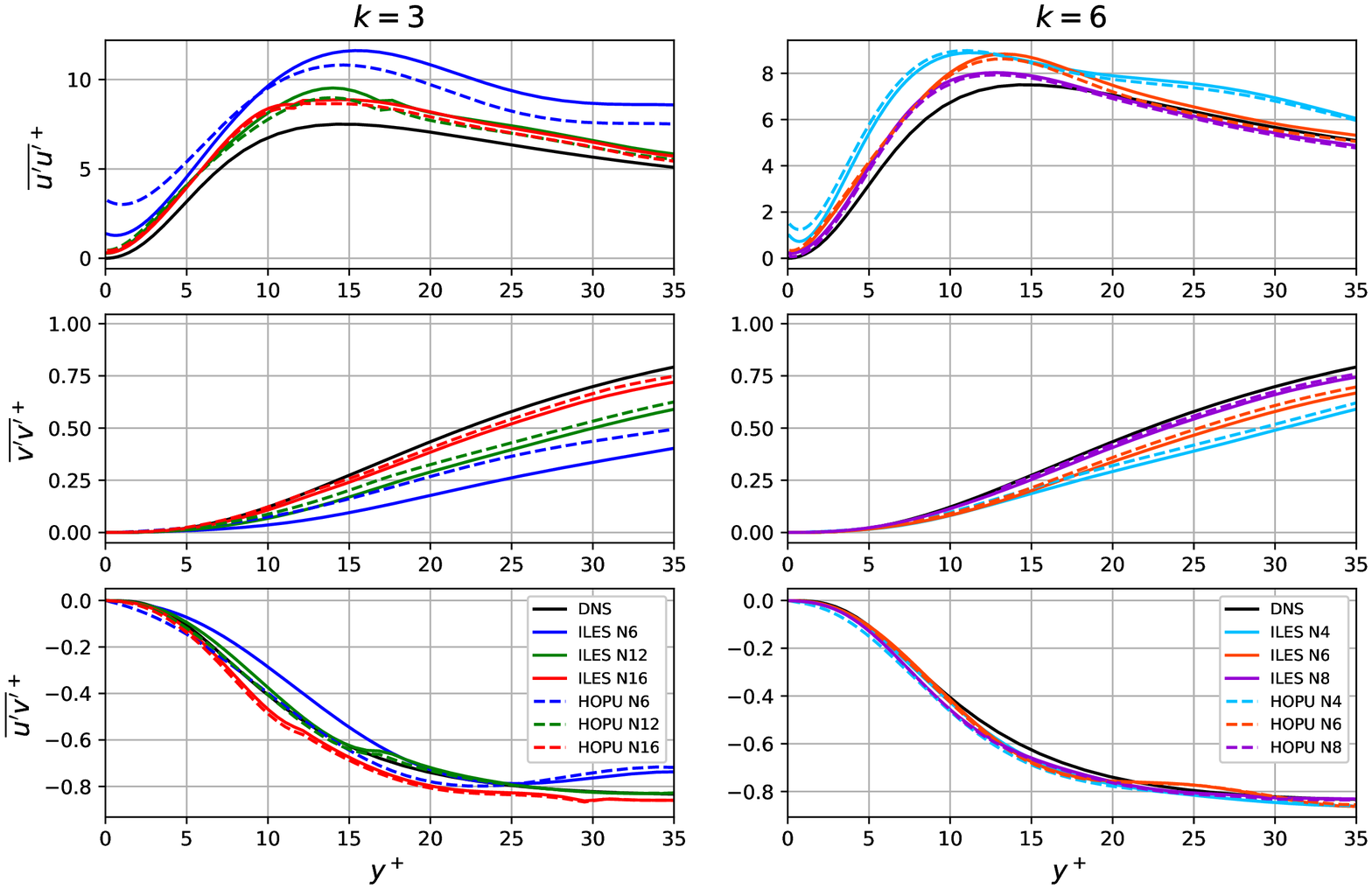}
    \caption{Normalized Reynolds stresses $\overline{u'u'}^+$, $\overline{v'v'}^+$ and $\overline{u'v'}^+$ over $y^+$ for the $p=6$ test case.}
    \label{fig:reynoldsstresses40}
\end{figure}

For the Reynolds stresses, no significant differences are observed
except slight improvements in the near-wall region. A more closed look
is given in figure \ref{fig:reynoldsstresses40}. The highly
under-resolved $N=6$ and $k=3$ HOPU case shows better agreement of the
different moments with the reference data in the proximity of the
wall. 

The relative error between the actual Reynolds number defined by the
mean bulk velocity $U_b = 1/(2 \pi)\int_{\partial \Omega |_{x=0}} u\:
d\underline{s}$, and the target Reynolds number, can be seen in table
\ref{table_channel_results}. Compared to the ILES and VMS simulations,
the HOPU variant gives significantly smaller errors in the range of
$\leqslant 1 \%$ for all resolutions. These results highlight the
observations made in figure \ref{fig:vms} and \ref{fig:hopu}. It is
important to bear in mind that the relative error based on the mean
velocity is only a rough error estimation of the approximation.

\begin{table}[h]
	\centering
	\begin{tabular}{|c|c|c|c|}
		\hline
		$N$/$k$ & 6/3 & 12/3 & 16/3\\ \hline
		ILES & 13.4\% & 3.1\% & 1.1\%\\
		VMS & 16.3\% & 4.7\% & 2.7\%\\
		HOPU & 0.7\% & 1.0\% & 0.6\%\\
		\hline
	\end{tabular}
	\vspace{3mm}
	\caption{Relative error of the target bulk velocity Reynolds number.}
	\label{table_channel_results}
\end{table}

\begin{figure}[h]
	\centering
    \includegraphics[width=1\textwidth]{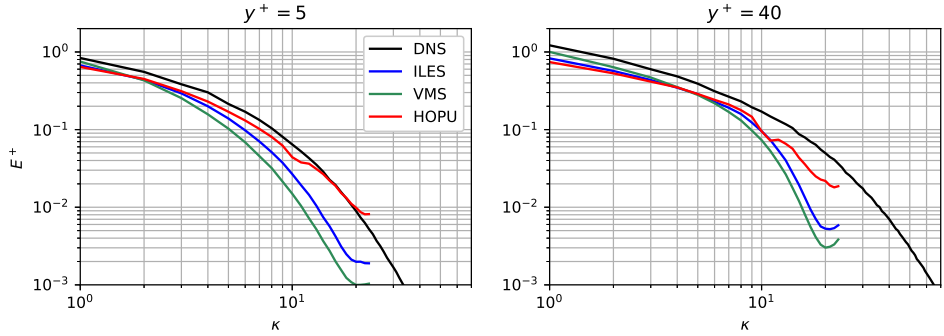}
    \caption{Normalized energy spectrum $E^+$ over the wave number $\kappa$ at two different planes, $y^+=5$ and $y^+=40$ respectively.}
    \label{fig:spectrum}
\end{figure}

\textbf{Comparison of the energy spectrum}:  In figure
\ref{fig:spectrum} we compare the methods by its energy spectrum
(obtained from the discrete Fourier transformation mentioned before).
The time-averaged normalized total energy spectrum $E^+$ over the wave
number $\kappa$ for the respective approaches is given. Its values are
calculated at two different planes normal to the $y$-direction. The
first plane is in the viscous sublayer at $y^+=5$ and the other one in
the buffer layer at $y^+=40$. The velocity data for the calculation of
the spectrum is provided from a simulation performed with a resolution
$N=12$/$k=3$.

In all cases we observe a good approximation of the spectrum in the
low-frequent range. As for higher values of $\kappa$, the energy
spectrum successively drops as it reaches the dissipative range. The
VMS method introduces additional dissipation to the mid to high range.
There, the energy spectrum begins to drop at lower wave numbers
compared to the ILES approach. Remarkably, the HOPU variant reduces
excessive damping and gives a much better prediction of the high wave
number range in comparison to the reference DNS data for $y^+=5$.

\section{Periodic hill flow}\label{periodichill}

The prediction of flow quantities from curved surfaces can be assased
by the well-known periodic hill flow test case originally proposed in
\cite{Mellen}. This type of flow induces several complex flow
phenomenons including reattachment, separation and strong interaction
with the outer flow. The transition of a boundary layer flow to a
separated shear layer can not be predicted by law of the wall and
standard model assumptions.

\begin{figure}[h]
	\centering
    \includegraphics[width=0.7\textwidth]{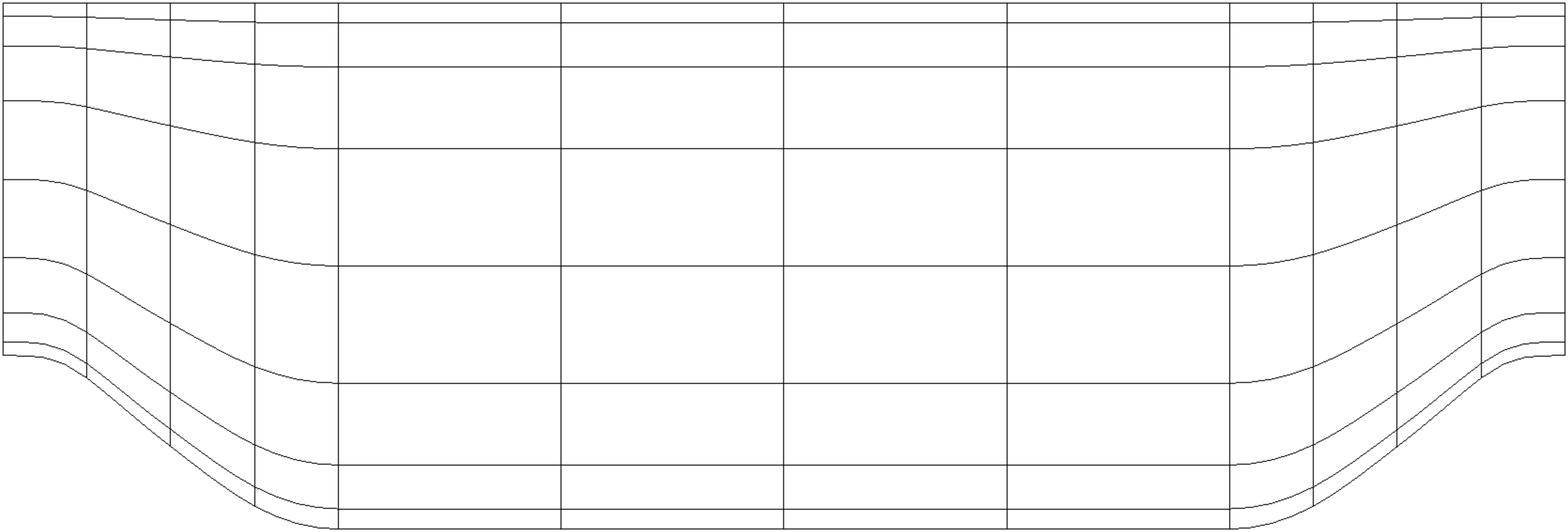}
    \caption{Curvilinear grid for the periodic hill problem.}
    \label{fig:grid}
\end{figure}

We use the same boundary conditions as for the channel flow. The grid
is composed of hexahedral elements that are first stretched with the
same stretching function (adapted to the height) as used in section
\ref{channel}, and then curved (using a third order polynomial
mapping) to fit the geometry see figure \ref{fig:grid}.
The dimensions of the domain $\Omega = (0,L_x) \times (0,L_y) \times
(0,L_z)$ are $L_x=9h$, $L_y=3.0335h$ and $L_z=4.5h$, where $h=1$
denotes the hill height. The characteristic flow is described by the
bulk velocity Reynolds number $Re = U_bh/\nu$, where the bulk velocity
is now given by 
\begin{equation*}
U_b = \frac{1}{(L_y-h)L_z}\int_{\partial \Omega |_{x=0}} u\: d\underline{s}.
\end{equation*}
In order to obtain a constant bulk velocity Reynolds number (in
contrast to a Reynolds number defined with respect to $\tau_w$ as for
the channel flow) the force on the right hand side is dynamically
controlled. For our results, we used $Re = 2800$ for all simulations.
Two different resolutions are used in this study, which can be seen in
table \ref{table_PH}. The same time step size is used as for the channel flow. The choice of polynomial order is $k=3$ for both
cases.

\begin{table}
	\centering
	\begin{tabular}{|c|c|c|c|c|c|}
		\hline
		& $k$ & $N_x$ & $N_y$ & $N_z$ & gDOF$\cdot 10^{-3}$ \\ \hline
		coarse & 3 & 12 & 8 & 6 & 43.2  \\
		fine & 3 & 24 & 16 & 12 & 331.0 \\
		\hline
	\end{tabular}
	\vspace{3mm}
	\caption{Simulation parameters for the periodic hill flow.}
	\label{table_PH}
\end{table}

After a statistically-steady state is achieved, the statistical
properties at different planes in x-direction are measured over a
given time period. The chosen planes are located at
$x:\{h,2h,3h,5h,8h\}$. The obtained solutions are compared with the
reference DNS data provided by \cite{Balakumar}.

\subsection{Comparison of results}\label{periodichill-compare}

\begin{figure}[h]
	\centering
    \includegraphics[width=1\textwidth]{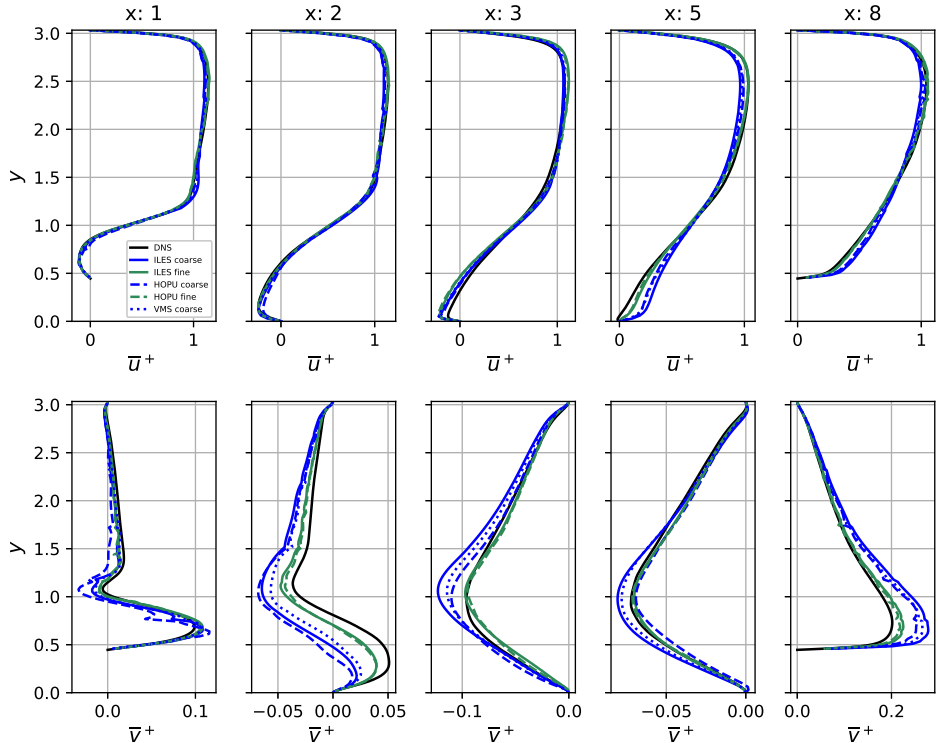}
    \caption{Normalized mean velocity $\overline{u}^+$ and $\overline{v}^+$ over $y$ at different planes in x-direction.}
    \label{fig:PH_u}
\end{figure}

Figure \ref{fig:PH_u} shows the profiles of normalized and averaged
velocity components from the different approaches at the considered
levels of resolution. The VMS approach uses the order $m=1$ for the
space $L_h$ and the adaptive HOPU uses the limits $\eta =
\{0.025,0.05,0.1\}$. It can be seen that all simulations are able to
reproduce  the streamwise velocity component $\overline{u}^+$
reasonably well. Notable differences are observed in the
approximations of $\overline{v}^+$. Naturally, the best match with the
reference data is obtained for the simulations on the fine grid. In
comparison of HOPU to no-model approach, we observe better aggrement
of $\overline{v}^+$ with HOPU at $x=\{3h,5h,8h\}$ for the coarse mesh.
The novel approach shows improved profiles in areas where flow
reattachment occurs. The results yielded by the ILES and VMS
approaches on the coarse grid are not fundamentally different.

\begin{figure}[h]
	\centering
    \includegraphics[width=0.8\textwidth]{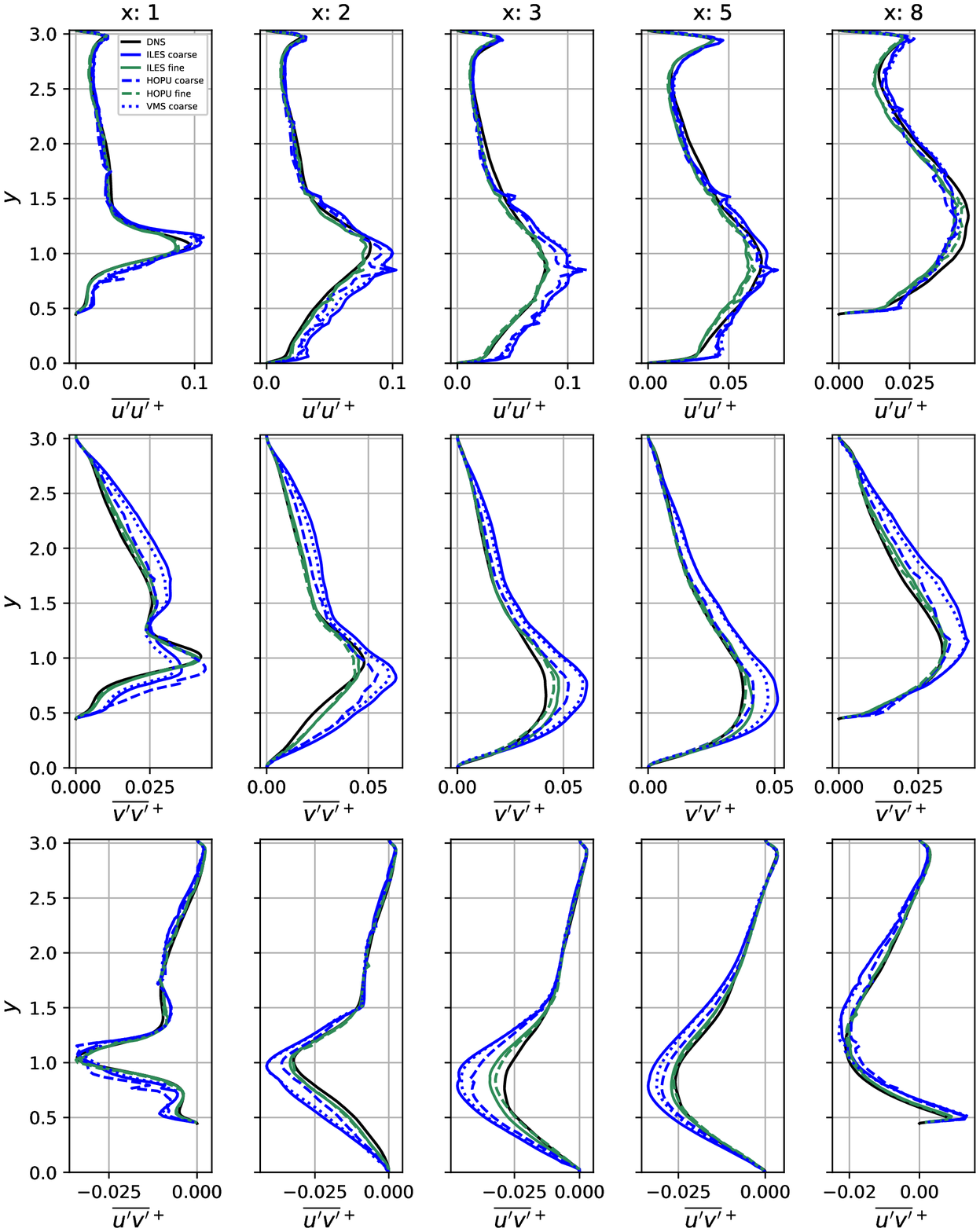}
    \caption{Normalized Reynolds stresses $\overline{u'u'}^+$, $\overline{v'v'}^+$ and $\overline{u'v'}^+$ over $y$ at different planes in x-direction.}
    \label{fig:PH_uu}
\end{figure}

The profiles of the Reynolds stresses can be seen in figure
\ref{fig:PH_uu}. As for the velocities, the solutions accomplished by
the finer grid shows overall good agreement in comparison with the
reference data. The results obtained by the coarse mesh mostly
overestimates the DNS profiles. For the Reynolds stresses
$\overline{v'v'}^+$ and $\overline{u'v'}^+$, the HOPU simulation
obtains significantly better profiles than the ILES approach. For the
VMS approach, only slight to no improvment is observed by the given
results.


\begin{figure}[h]
	\centering
    \includegraphics[width=0.8\textwidth]{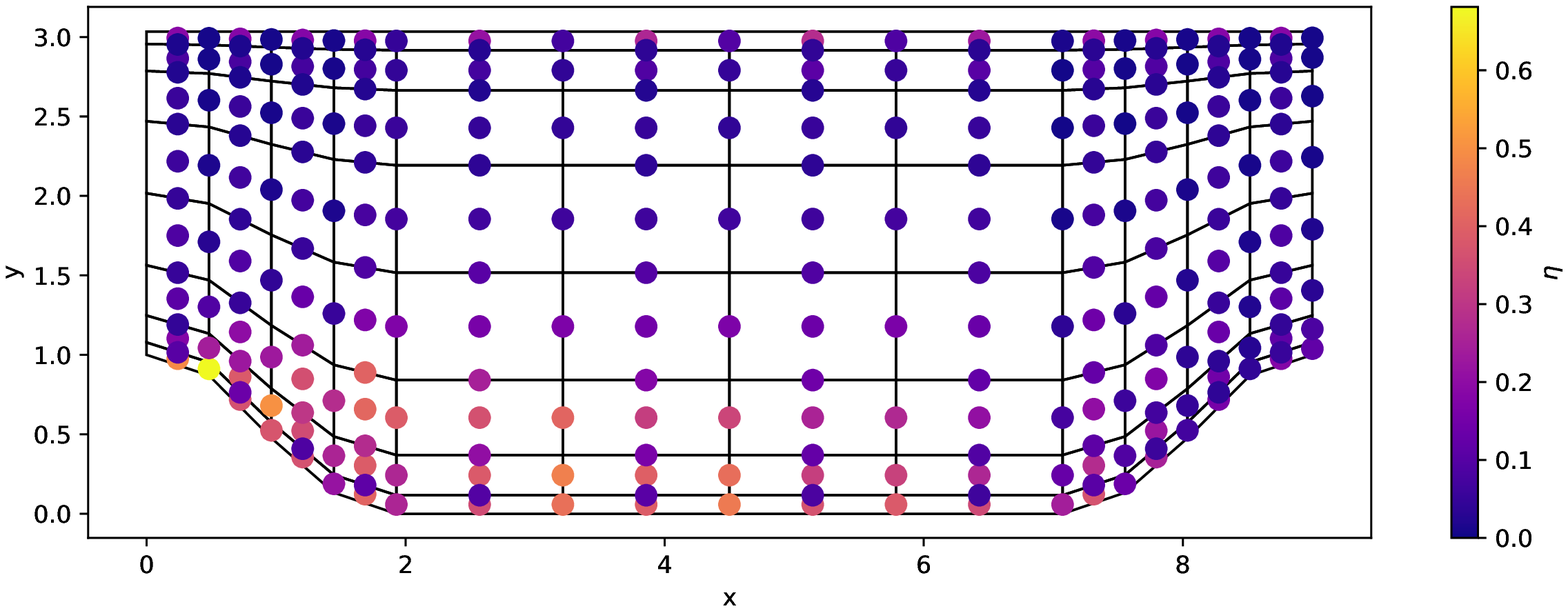}
    \caption{Average absolute value jump $\eta$ for the periodic hill flow problem.}
    \label{fig:eta_PH}
\end{figure}

The averaged absolute value jump calculated from the coarse ILES
simulation is given in figure \ref{fig:eta_PH}. High values of $\eta$
are observed in the proximity of the lower wall and especially in the
downstream region of the hill. Interestingly, the peak value seems to
be located in the immediate vicinity of the separation point. This
figure gives a rough representation of the areas, which are mostly
affected by the HOPU approach.

\section{Conclusion}\label{summary}

The capacity of the MCS approach to capture structure-resolving
turbulent flow in a wall-bounded configuration has been assessed for
two different test cases. In addition to the use of a no-model and
explicit VMS method, an entirely new technique called high-order
projected upwinding, has been introduced to further improve the
in-built dissipation capabilites. 

In view of the results, we can conclude that the use of a SGS model
does not appear to be fundamentally different to the no-model
approach. The WALE model incoporated with the VMS approach seems to
have an inferior implification on the DG discretization. Furthermore,
artificial dissipation introduced by the model can harm the implicit
dissipation mechanism and may lead to slightly worse results.\\
The new high-order projected upwind method significantly improves the
results of the standard ILES method for $k=3$ in many areas,
especially in the case of the turbulent channel flow. The problem of
excessive numerical dissipation introduced by the convection
stabilization term in the near-wall region can be avoided by HOPU. It
allows for less penalized turbulent transport in this area and overall
improved impact on the resolved structures. Results obtained by the
discretization using $k=6$ have shown to be not affected considerably
by HOPU.

We conclude that the adaptive local order HOPU can be interpreted as a
wall model approach for standard DG ILES discretization. However, it
is still unclear whether using the absolute value jump as an indicator
is the best practice and additional research regarding the adaptivity
is needed. However, even in flows where separation appears as in the
periodic hill case, HOPU showed an improvement where reattachment
appears. By these insights, the novel HOPU method applied to complex
turbulent flow problems has to be further investigated in future work.

\section{Acknowledgments}
This research was funded in part, by the Austrian Science Fund (FWF)
[F65-P10 and P35931-N]. For the purpose of open access, the author has
applied a CC BY public copyright licence to any Author Accepted
Manuscript version arising from this submission.

\clearpage

\printbibliography

\end{document}